\documentclass[12pt,preprint,url]{aastex}
\usepackage{amsbsy}
\usepackage{amsmath}
\newcommand{\be}{\begin{equation}}
\newcommand{\ee}{\end{equation}}

\def\ltsima{$\; \buildrel < \over \sim \;$}

\def\lsim{\lower.5ex\hbox{\ltsima}}
\def\gtsima{$\; \buildrel > \over \sim \;$}
\def\gsim{\lower.5ex\hbox{\gtsima}}

\shorttitle{Pulsar timing noise from superfluid turbulence}
\shortauthors{Melatos and Link}

\begin{document}
\title{Pulsar timing noise from superfluid turbulence}

\author{Andrew Melatos}
\affil{School of Physics, University of Melbourne,
Parkville, VIC 3010, Australia}
\email{amelatos@unimelb.edu.au}
\and
\author{Bennett Link} 
\affil{Department of Physics, Montana State University,
Bozeman, MT 59717--3840, USA}
\email{link@physics.montana.edu}


\begin{abstract}
\noindent 
Shear-driven turbulence in the superfluid interior of a neutron star
exerts a fluctuating torque on the rigid crust, 
causing the rotational phase to walk randomly.
The phase fluctuation spectrum is calculated
analytically for incompressible Kolmogorov turbulence
and is found to be red;
the half-power point is set by the
observed spin-down rate, the crust-superfluid lag,
and the dynamical response time of the superfluid.
Preliminary limits are placed on the latter quantities 
using selected time- and frequency-domain data.
It is found that measurements of the normalization and slope 
of the power spectrum
are reproduced for reasonable choices of the turbulence parameters.
The results point preferentially to the neutron star interior
containing a turbulent superfluid rather than a turbulent Navier-Stokes fluid.
The implications for gravitational wave detection by pulsar timing arrays
are discussed briefly.
\end{abstract}


\section{Introduction 
 \label{sec:tmg1}}
Timing noise is a type of rotational irregularity
observed in all isolated radio pulsars,
in which pulse arrival times wander stochastically about the
fitted ephemeris
\citep{boy72,cor80,cor85,dal95,bay99,hob10,sha10}.
It is characterized as a random walk 
in the rotational phase, angular velocity, or torque;
the Fourier spectrum is always red,
implying a process autocorrelated on a time-scale of hours to years
\citep{cor80}.
Timing noise has been attributed to various mechanisms:
microjumps akin to small glitches
\citep{cor85,jan06,mel08},
recovery from unseen glitches
\citep{joh99},
quasiperiodic changes in magnetospheric structure
\citep{lyn10},
variable coupling between the crust and liquid interior
\citep{alp86,jon90},
stochastic variations in the star's shape
\citep{cor93},
and fluctuations in the spin-down torque
\citep{che87b,che87a,ura06}.
The most recent comprehensive survey of timing irregularities
in 366 pulsars over time-scales longer than a decade
found that a low-frequency noise process cannot explain
all the observations on its own
\citep{hob10},
suggesting that more than one physical mechanism contributes.
In some pulsars, the pulse times-of-arrival are correlated
over weeks with the distinctive signature of a relaxation process,
e.g.\ damping of internal differential rotation
\citep{pri12}.

Recent research is lending growing support to the hypothesis,
propounded originally by \citet{gre70}, that the superfluid
interior of a neutron star is turbulent. The turbulence takes
one of two forms: macroscopic, Kolmogorov-like eddies driven by 
crust-core shear at high Reynolds numbers, 
possibly involving unstable structures like Stewartson layers
\citep{per05,per06a,per06b,mel07,per09,mel12};
and microscopic tangles of self-regenerating, reconnecting, quantized vorticity, 
driven by Kelvin-wave instabilties like those seen in terrestrial superfluids
\citep{per06b,and07}
or dissipative instabilities arising from perfect or imperfect pinning
in the inner crust or outer core
\citep{lin12b,lin12a}.
It has been suggested that these forms of turbulence
develop in a sustained manner and contribute to
the stochastic spin variations observed in radio pulsars
\citep{gre70,per07,lin12a}.

In this paper, we calculate from first principles the torque statistics 
and phase wandering produced by superfluid turbulence. 
The effect sets a timing noise floor,
on top of which other processes like magnetospheric state changes
add their contributions.
In \S\ref{sec:tmg2}, we calculate the autocorrelation function
for angular momentum fluctuations 
in the context of an idealized neutron star model.
In \S\ref{sec:tmg3}, we calculate the 
power spectral density of the phase residuals
and present convenient analytic formulas for the
roll-over frequency and zero-frequency normalization.
In \S\ref{sec:tmg4}, we compare the predicted spectrum with timing data
from a few representative objects and show how to place
limits on quantities of fundamental physical importance,
like the dynamical response time of the neutron superfluid and the moment
of inertia of the stellar crust,
mindful that a low-frequency noise process cannot explain 
all the irregularities observed 
\citep{hob10},
and that a comprehensive comparison with more data must still be done.
The calculations are closely related to techniques developed
to compute the stochastic gravitational radiation emitted by
a turbulent neutron star 
\citep{mel10}
and by phase transitions in the early Universe
\citep{kos02,gog07}.

\section{Turbulent torque
 \label{sec:tmg2}}

\subsection{Idealized neutron star model
 \label{sec:tmg2a}}
We start by considering an idealized model of a neutron star
as two coupled subsystems. The first subsystem is the rigid crust
and charged electron-proton fluid, which we assume
are locked together by magnetic stresses and hence corotate
\citep{alp84a,rud98}.
\footnote{
Crust-core corotation is not guaranteed.
The hydromagnetic coupling is subtle 
\citep{mel12};
it is weakened by buoyancy
\citep{men98},
type I superconductivity
\citep{has12},
thermally activated vortex creep
\citep{lin13},
and in nontrivial magnetic geometries
\citep{eas79b}.
}
The second subsystem is the inviscid neutron condensate. 
Our interest is in the scenario where the condensate is turbulent, 
driven by one or more of the processes referenced
in \S\ref{sec:tmg1}.
For simplicity, we assume that the mass density $\rho$ of the star
is uniform.

The two subsystems couple through friction. The exact nature of the friction
is unimportant for this paper, but we now describe some relevant processes
to give physical context. In the outer core, the dominant contribution
to friction appears to be the scattering of electrons off vortices in the
neutron condensate, whose cores are magnetized by entrainment of the
neutron and proton mass currents
\citep{alp84a}.
This interaction could be modified significantly by vortex clustering
\citep{sed95a,sed95b}.
The protons are expected to form a type II superconductor, 
with the magnetic flux confined in flux tubes, which are frozen to
the highly conducting, charged plasma.
The vortices of the neutron superfluid pin to the flux tubes,
primarily through a magnetic interaction, with pinning energies
as high as $\sim 10^2\,{\rm MeV}$ per vortex-flux-tube junction
\citep{sri90,jon91,men91a,cha92,rud98,gla11,lin12b}.
Pinning partly decouples the neutron and charged fluids,
increasing the coupling time-scale
\citep{lin13}
and sustaining an angular velocity difference,
as the charged component of the star is spun down by the external,
electromagnetic torque.
While pinning energies remain rather uncertain,
the conclusion that vortices pin to outer-core flux tubes 
appears to be increasingly likely (provided the outer core
is indeed a type II superconductor);
this point is discussed further in \S\ref{sec:tmg2aa}.

For simplicity, we assume that angular momentum transport through
the neutron condensate occurs instantaneously.
This approximation is good, even if angular momentum is stored
temporarily in a third subsystem,
e.g.\ Kelvin waves propagating
along superfluid vortices pinned to the inner crust,
or hydromagnetic-inertial and cyclotron-vortex waves
propagating through the charged fluid
\citep{eas79a,men98,mel12};
the associated wave-crossing time-scales are still fast
($\lesssim 10^2\,{\rm s}$) in a typical star.

We suggest as a useful mechanical analogy a boiling pot 
of water on a frictionless stove.
As the water (the turbulent condensate) boils, 
the total angular momentum of the water fluctuates,
and a stochastic torque is applied to the pot (the charged component).
This paper is concerned with analysis of the stochastic torque
in the context of a neutron star.

In what follows, we regard the turbulent condensate as
driving the crust, not vice versa; that is,
the angular random walk executed by the crust does not feed back 
to modify the turbulence,
at least on observational time-scales of decades. 
We justify this approximation quantitatively 
a posteriori in \S\ref{sec:tmg4a}.

\subsection{Turbulent condensate
 \label{sec:tmg2aa}}
The physics of the turbulent condensate has been examined
previously by many authors
\citep{per05,per06a,and07,mel10,lin12b,lin12a}.
Uncertainties remain.
Large-scale simulations in the nonlinear regime have been performed
for some driving mechanisms,
but even so the limited dynamic range means that important physics 
is not always captured.
Here we review the main possibilities briefly,
emphasizing those aspects that motivate the idealized model
developed in \S\ref{sec:tmg2} and \S\ref{sec:tmg3}.
In essence, the model postulates the existence of turbulence with
Kolmogorov-like statistics in some fluid component
that couples to the crust.
There are many ways to realize this scenario,
and we now describe some of them.

Turbulence driven by vortex instabilities can arise in the core
and/or the inner crust.
In the core, where the vortices may be pinned to flux tubes,
imperfect pinning destabilizes the vortex lattice;
the source of free energy is the differential rotation
between core neutrons and the proton-electron fluid,
which locks magnetically to the crust
\citep{lin12b}.
In the absence of turbulence, 
the angular velocity lag is $\sim 0.1\,{\rm rad\,s^{-1}}$;
in its presence, the lag and hence the steady-state injected power
per unit enthalpy
(labeled $\varepsilon$ in \S\ref{sec:tmg2b})
remain unknown,
because the nonlinear saturation time-scale has not yet been calculated.
Pinning-driven vortex instabilities can also occur in the inner crust
\citep{lin12a},
with pinning by flux tubes replaced by pinning at nuclear lattice sites.
The two scenarios are essentially identical with regard to the calculations
in this paper, the main difference being the inertia carried
by the turbulent condensate (e.g.\ core neutrons versus inner crust neutrons)
and associated entrained components.

Turbulence driven by meridional circulation (e.g.\ unstable Ekman pumping
at high Reynolds number) can also occur in the core and/or inner crust.
Again, the angular velocity difference and steady-state power 
have not been calculated self-consistently in the literature;
the angular velocity of the outer crust is specified by fiat
in simulations,
without adjusting for the back-reaction torque from the viscous
proton-electron component
\citep{per05,per06a}.
Nevertheless, for realistic neutron star parameters, 
the differential velocity projected along the rotation axis 
arising from spin-down-powered Ekman circulation greatly exceeds the 
Donnelly-Glaberson instability threshold
\citep{per05,and07}
in large parts of the core and inner crust,
generating islands of tangled vorticity and patchy mutual friction
\citep{per06a}.
\footnote{
The mutual friction force is ``patchy'' in the sense that it takes
different forms locally.
Specifically, it is isotropic where the vortices are tangled,
anisotropic in a rectilinear vortex array, and $\sim 10^3$ times weaker in the 
former configuration than in the latter for typical neutron star parameters
\citep{per06a}.
}
Liquid helium experiments show that 
turbulent velocity spectra in a superfluid are Kolmogorov-like
(over two decades in wave number)
in various grid, wake, and ``chunk'' flows
\citep{sal10}.
The chief theoretical input, $\varepsilon$,
is left as a free parameter,
to be constrained by pulsar timing data,
although it can also be estimated robustly
from the spin-down rate and angular velocity difference
by an energy balance argument.

If pinning is strong (due to flux tubes in the core or nuclear
lattice sites in the inner crust), 
the characteristic time-scale over which turbulent structures change, 
$\tau_{\rm turb}$, is longer than the usual eddy turnover time-scale,
$\tau_{\rm eddy}$;
i.e.\
the dynamical time-scale for tangled vorticity is longer
in the condensate than in any unpinned components.
The ratio 
\begin{equation}
 \gamma = \tau_{\rm eddy}/\tau_{\rm turb} \leq 1~,
\end{equation}
which we call the `pinned turbulence parameter',
tends to zero for perfect pinning.
In general, pinning is imperfect;
vortices move slowly via
thermally activated vortex creep
\citep{alp84b,lin93,sid09,lin13}
or sporadic avalanches
\citep{war08,mel09,war11,war12,war12b},
with the creep or avalanche rate adjusting in response to 
the local Magnus force to set $\gamma$.
We are mostly interested in the regime $\gamma \ll 1$,
corresponding to the unusual situation of ``pinned turbulence''
in a superfluid;
by contrast, in standard Navier-Stokes turbulence, one has $\gamma=1$.
The factor $\gamma$ has not yet been measured in laboratory experiments 
or calculated self-consistently;
indeed, it is likely to be a function of the turbulent state itself.
In the theory developed below, however,
it always appears in the combination $\varepsilon^{1/3} \gamma$,
avoiding the need to introduce an extra degree of freedom.

In this paper, we assume that the decelerating crust 
comes into dynamical equilibrium with the Kolmogorov cascade in the condensate.
The existence of a continuously driven, statistically steady, turbulent state 
(driven here by electromagnetic spin down) 
has been confirmed experimentally in Navier-Stokes and superfluid turbulence
\citep{sag08,sal10}.
Stratification can quench the turbulence intermittently under
certain conditions
\citep{chu12,las13},
but such quenching is incompletely understood even in terrestrial contexts
and falls outside the scope of this paper.

\subsection{Angular momentum fluctuations
 \label{sec:tmg2b}}
In a suitably time-averaged corotating frame, 
the instantaneous total angular momentum 
of the turbulent condensate is
\begin{equation}
 \delta J_i(t) 
 = 
 \epsilon_{ijk}
 \int d^3{\bf x} \, 
 \rho({\bf x},t) x_j v_k({\bf x},t)~,
\label{eq:tmg1}
\end{equation}
where ${\bf v}({\bf x},t)$ is the superfluid velocity field in the
corotating frame,
and the integral is taken over the volume $V$ of the star
(a sphere of radius $R$ in our idealized model).
As the system is isolated,
$\delta {\bf J}(t)$ averages to zero over intervals much longer than
the eddy turnover time-scale for the largest eddies,
but the mean-square fluctuations are not zero.
From (\ref{eq:tmg1}), we form the autocorrelation function
\begin{eqnarray}
 G_{ij}(\tau)
 & = &
 \langle
  \delta J_i(t) \delta J_j(t')^\ast
 \rangle
\label{eq:tmg2}
\\
 & = &
 \epsilon_{ilm} \epsilon_{jpq} \rho^2
 \int_0^R dr \, r^3
 \int_0^R dr' \, (r')^3
 \int d^2{\bf \hat x} \int d^2{\bf \hat x}'
 \int \frac{d^3{\bf k}}{(2\pi)^3} \int \frac{d^3{\bf k}'}{(2\pi)^3}
\nonumber 
\\
 & & \times
 \exp(i{\bf k}\cdot{\bf x}) \exp(-i{\bf k}'\cdot{\bf x}')
 {\hat x}_l {\hat x}_p'
 \langle v_m({\bf k},t) v_q({\bf k}',t')^\ast \rangle
\label{eq:tmg3}
\end{eqnarray}
with $\tau=t'-t$.
Angle brackets denote an ensemble average
over multiple realizations of the turbulent flow.
We pass from (\ref{eq:tmg2}) to (\ref{eq:tmg3}) by expanding 
${\bf v}({\bf x},t)$ as a sum of spatial Fourier modes ${\bf v}({\bf k},t)$.

In isotropic Kolmogorov turbulence, the velocity autocorrelation function
for Fourier modes at wave numbers ${\bf k}$ and ${\bf k}'$ takes the form
\citep{kra59}
\begin{equation}
 \langle v_m({\bf k},t) v_q({\bf k}',t')^\ast \rangle
 =
 V(2\pi)^3 (\delta_{mq} - {\hat k}_m {\hat k}_q)
 P(k) \exp[- \eta(k) | \tau | ] 
 \delta({\bf k}-{\bf k}')~,
\label{eq:tmg4}
\end{equation}
with
\begin{equation}
 \eta(k) = (2\pi)^{-1/2} \varepsilon^{1/3} k^{2/3} \gamma
\label{eq:tmg5}
\end{equation}
and
\begin{equation}
 P(k) = V^{-1} \pi^2 \varepsilon^{2/3} k^{-11/3} \gamma^2~.
\label{eq:tmg6}
\end{equation}
The power spectrum $P(k)$
extends as a power law from the outer (stirring) scale
$k_{\rm s} = 2\pi/R$ to the inner (dissipation) scale
$k_{\rm d} = (8\varepsilon / 27\nu^3)^{1/4}$.
The kinematic viscosity $\nu$ in a superfluid arises from a combination
of particle collisions moderated by transverse Landau damping
\citep{sht08}
and a Kelvin wave cascade on vortex lines;
the latter channel is especially prominent under turbulent conditions
in helium
\citep{elt07,wal07,van12}.
The value of $\nu$ inside a neutron star is still poorly known,
but the theory in this paper is insensitive to it
with respect to timing noise observables.
In (\ref{eq:tmg5}) and (\ref{eq:tmg6}), $\varepsilon$ is the
power injected into the turbulence
per unit enthalpy 
(units: ${\rm cm^2 s^{-3}}$; see \S\ref{sec:tmg2aa}),
and $\gamma$ enters $P(k)$ quadratically,
because the turbulent velocity at wave number $k$ is proportional to $\gamma$
due to pinning (see \S\ref{sec:tmg2aa}).
Equation (\ref{eq:tmg4}) applies regardless of the specific driver 
and dissipation process, as long as isotropy is maintained.
Many factors intervene in reality to destroy isotropy,
e.g.\ coherent structures like hairpin vortices or wall-wake flows,
anomalous Reynolds stresses, and compositional stratification.
They fall outside the scope of this paper but
are studied extensively in wind tunnel and grid turbulence 
experiments,
as noted in \S{2.2} and \S{3.3} of \citet{mel10} and references therein.

Two-point velocity fluctuations decorrelate exponentially
with $\tau$ according to (\ref{eq:tmg4}).
There is some latitude inherent in the functional form:
laboratory data and numerical simulations variously point to 
an exponential or Gaussian cut-off in Navier-Stokes turbulence
\citep{com71,don08},
while analogous measurements in a superfluid have never been done
\citep{sal10},
and the role played by long-duration intermittency is still unclear
\citep{mer12,zra12}.
In the absence of definitive experiments,
we adopt the exponential form here,
anticipating the empirical finding that the spectrum of pulsar timing noise
is observed to be red
\citep{hob10}
with a power-law high-frequency tail.
\footnote{
Intermittency leads to a flatter tail than the Gaussian
predicted by Kraichnan's random sweep process
\citep{che89,kos02}.
}
The decorrelation time-scale $\eta(k)^{-1}$ 
and kinetic energy per unit wave number $k^2 P(k)$
are proportional to
$\gamma^{-1}$
and
$\gamma^{2}$
respectively,
as discussed in \S\ref{sec:tmg2aa};
in general, $\gamma$ itself may also be a function of $k$.
In the limit $\gamma \rightarrow 0$, the turbulence is quenched.
As superfluid turbulence remains poorly understood
\citep{sal10},
especially in rotating systems,
it is hard to compute $\gamma$ reliably from first principles.
Instead, as foreshadowed in \S\ref{sec:tmg2aa},
we keep it as a model parameter 
(always appearing in the combination $\varepsilon^{1/3}\gamma$)
and explain in \S\ref{sec:tmg4}
how pulsar timing noise measurements constrain it.

To simplify (\ref{eq:tmg3}), we follow the procedure laid out
in {\S}3.2 in \citet{mel10}:
integrate with respect to ${\bf k}'$ over the delta function,
expand the plane wave factors in spherical harmonics,
then integrate over 
${\bf \hat x}$, ${\bf \hat x}'$, and ${\bf \hat k}$.
Details of the algebra are given in Appendix \ref{sec:tmgappa}. 
The remaining integral over $x=kR$ involves
squares of rapidly oscillating spherical Bessel functions
in the integrand.
Averaging over many cycles of the fast oscillation, we obtain,
to a good approximation,
\begin{equation}
 G_{ij}(\tau)
 =
 \frac{8\pi^2}{15}
 \rho^2 \varepsilon^{2/3} R^{26/3} \gamma^2 \delta_{ij}
 \int_{k_{\rm s}R}^{k_{\rm d}R} dx\,
 x^{-29/3} ( x^4 + 3 x^2 + 9 )
 \exp\left[
 - x^{2/3} \eta(R^{-1}) | \tau |
 \right]~.
\label{eq:tmg7}
\end{equation}
The integral in (\ref{eq:tmg7}) is dominated by its lower terminal.
Its maximum value at zero lag is
\begin{equation}
 G_{ij}(0) = 2.2\times 10^{-4}
 \rho^2 \varepsilon^{2/3} R^{26/3} \gamma^2 \delta_{ij}~.
\label{eq:tmg7a}
\end{equation}
One also finds
\begin{equation}
 G_{ij}( | \tau | = 0.44 \varepsilon^{-1/3} R^{2/3} \gamma^{-1} )
 =0.50 G_{ij}(0)
\label{eq:tmg7b}
\end{equation}
at the half-power point.

\section{Phase residuals
 \label{sec:tmg3}}
Timing of radio pulsars gives the
autocorrelation function of the phase residuals $\delta\phi(t)$, i.e.\
\begin{equation}
 \langle
 \delta\phi(t) \delta\phi (t')
 \rangle
 = 
 \frac{1}{4\pi^2}
 \int_0^t dt'' \int_0^{t'} dt'''
 \langle
 \delta\Omega_z(t'') \delta\Omega_z (t''')
 \rangle~,
\label{eq:tmg9a}
\end{equation}
where $\delta{\bf \Omega}(t)$ denotes the angular velocity in the
time-averaged corotating frame introduced in \S\ref{sec:tmg2b}.
We choose $\delta{\bf \Omega}(t)$ to lie along the $z$ axis and
assume that its orientation remains fixed in the observer's frame;
there is no observational evidence for any secular drift of 
the pulse profile 
(e.g.\ relative height or separation of conal components) 
or linear polarization swing in any of the objects
monitored by \citet{hob10} over decades.
Approximating the star as spherical,
and assuming instantaneous angular momentum transfer 
via magnetic coupling
between the turbulent condensate, charged electron-proton fluid,
and crust, as in \S\ref{sec:tmg2a},
we write
$\delta {\bf J} = - I_{\rm c} \delta {\bf \Omega}$ 
and hence
$\langle \delta\Omega_i(t) \delta\Omega_j(t') \rangle = 
 G_{ij}(\tau) / I_{\rm c}^2$.
Here, $I_{\rm c}$ is the moment of inertia of the solid crust plus
rigidly corotating charged fluid
plus entrained neutrons
\citep{cha12},
making up an effective noncondensate fraction 
\begin{equation}
 \lambda=I_{\rm c}/I_0
\end{equation} 
of the total moment of inertia $I_0=8\pi R^5 \rho / 15$
(notionally, if the whole star were a rigid body).

The phase residuals $\delta\phi(t)$ satisfying (\ref{eq:tmg9a})
constitute a nonstationary time series;
the left-hand side of (\ref{eq:tmg9a}) is a function of $t$ and $t'$
independently, not solely through the combination $\tau$,
unlike $G_{ij}(\tau)$.
For $t' > t$, it is helpful to change integration variables 
from $(t'', t''')$ to $(t'',t'''-t'')$ and use the property
$G_{ij}(-\tau)=G_{ij}(\tau)$ to evaluate (\ref{eq:tmg9a}):
\begin{eqnarray}
 \langle
 \delta\phi(t) \delta\phi (t')
 \rangle
 & = & 
 \frac{1}{4\pi^2 I_{\rm c}^2}
 \left[
  \int_{0}^t d\tau \, (t - \tau)
  +
  t \int_{0}^{t'-t} d\tau \,
  +
  \int_{t'-t}^{t'} d\tau \, (t' - \tau)
 \right]
 G_{zz}(\tau)
 \label{eq:tmg9b}
 \\
 & = &
 \frac{15}{16\pi\lambda^2}
 \int_{2\pi}^\infty dx \, x^{-11}
 (x^4 + 3x^2 + 9)
 \left[
  - 1
  + 2 \eta(R^{-1}) x^{2/3} t
  \phantom{ e^{-\eta(R^{-1}) x^{2/3} t} }
 \right.
 \nonumber
 \\
 & & 
 \left.
  +
  e^{-\eta(R^{-1}) x^{2/3} t'}
  -
  e^{-\eta(R^{-1}) x^{2/3} | \tau |}
  +
  e^{-\eta(R^{-1}) x^{2/3} t}
 \right]~.
 \label{eq:tmg9c}
\end{eqnarray}
The result for $t>t'$ is the same but with $t$ and $t'$ swapped.
The first two terms in square brackets in (\ref{eq:tmg9c}) describe a
secular drift,
which grows slowly over many multiples of the decorrelation time-scale
and is absorbed into the best-fit ephemeris measured by absolute pulse numbering.
Physically the drift parallels the monotonic increase in mean-square
displacement with time in standard Brownian motion.
\footnote{
The two secular terms combine with the other three terms to give zero drift
as $t$, $t' \rightarrow 0$, as required.
}
The third and fifth terms are transients which decay quickly
on the time-scale $\approx \eta(R^{-1})^{-1}$ and are unimportant
away from the origin $t=t'=0$.
\footnote{
The observed statistics depend on the time origin of the measurements,
because the random walk prior to $t=t'=0$ adulterates the random walk
at $t$, $t' > 0$ by randomizing $\delta\phi(t)$ and its derivatives
at $t=0$,
as proved in {\S}3c of \citet{cor80b}.
}
The fourth term, which persists for all $t$, $t' \geq 0$, 
describes the stationary piece of the $\delta\phi(t)$ time series.
It contains the underlying phase residual statistics 
after the ephemeris is subtracted,
and its spectrum is red,
because the fluctuating torque from superfluid turbulence is correlated temporally.
Letting $\Phi(\tau)$ denote the absolute value of the contribution to
$\langle \delta\phi(t) \delta\phi(t') \rangle$ from the fourth term,
we Fourier transform $\Phi(\tau)$ to obtain the power spectral density,
\begin{eqnarray}
 \Phi(f)
 & = &
 \int_{-\infty}^\infty d\tau \, \exp(2\pi i f \tau) \Phi(\tau)
 \label{eq:tmg11}
 \\
 & = &
 \frac{15}{8\pi\lambda^2 \eta(R^{-1})}
 \int_{2\pi}^\infty
 dx\,
 x^{-31/3} (x^4+3x^2+9)
  \{ [ 2\pi f / \eta(R^{-1}) ]^{2} + x^{4/3} \}^{-1},
 \label{eq:tmg12}
\end{eqnarray}
where $f$ is the Fourier frequency.
The spectrum (\ref{eq:tmg12}) can be compared directly against
radio telescope timing data.
It is red, as predicted, with $\Phi(f) \propto f^{-2}$ at large $f$.

We can measure the fundamental parameters $\lambda$ and 
$\varepsilon^{1/3} R^{-2/3} \gamma$ by measuring $\Phi(f)$
in individual pulsars.
If an object is monitored long enough,
one can eventually see $\Phi(f)$ roll over,
read off the half-power point
$[f_{1/2},\Phi(f_{1/2})=0.5 \Phi(f=0)]$,
and hence find
\begin{equation}
 \lambda = 5.2 \times 10^{-4} [ f_{1/2} \Phi(f=0) ]^{-1/2}~,
 \label{eq:tmg13}
\end{equation}
\begin{equation}
 \varepsilon^{1/3} R^{-2/3} \gamma = 4.2 f_{1/2}~.
 \label{eq:tmg14}
\end{equation}
On the other hand,
if $\Phi(f)$ rolls over at a frequency below the observed range,
one can fit the $f^{-2}$ tail directly
and read off
\begin{equation}
 \varepsilon^{1/3} R^{-2/3} \lambda^{-2} \gamma
 =
 1.5 \times 10^7 f^2 \Phi(f)~.
 \label{eq:tmg15}
\end{equation}
Equations (\ref{eq:tmg13})--(\ref{eq:tmg15}) are our chief results.
They relate the parameters of the turbulence to the observed
power spectrum.

\section{Comparison with observations
 \label{sec:tmg4}}
We now undertake some preliminary comparisons between theory and data
to lay the groundwork for more comprehensive population studies 
in the future.
In \S\ref{sec:tmg4a},
we verify that the theory predicts roughly the correct
normalization and shape of $\Phi(f)$
for two representative objects with well-measured spectra,
given sensible choices of the underlying physical variables. 
This is just a rudimentary consistency check;
the constraints thereby derived on $\lambda$ and
$\varepsilon^{1/3} R^{-2/3}\gamma$ are indicative only;
the ultimate goal is to place unified constraints 
on these quantities across the pulsar population.
In \S\ref{sec:tmg4b}, we begin the latter task by examining
time-domain, root-mean-square measures of timing noise 
in a sample of 366 objects,
most of which do not yet have $\Phi(f)$ measured.
We find that the theory predicts $\Omega$ and $\dot{\Omega}$ scalings
in accord with the data.
The residual scatter may contain clues about how $\gamma$
(and hence the physics of pinning) varies across the pulsar population.
It deserves further study.
We caution that coefficients like $\gamma$
are governed by nonequilibrium transport processes,
so the existence of a simple, one-parameter family of models 
(indexed by stellar mass or temperature, for example) 
is not guaranteed.

\subsection{Power spectral density
 \label{sec:tmg4a}}
\begin{figure}
\begin{center}
\includegraphics[width=14cm,angle=0]{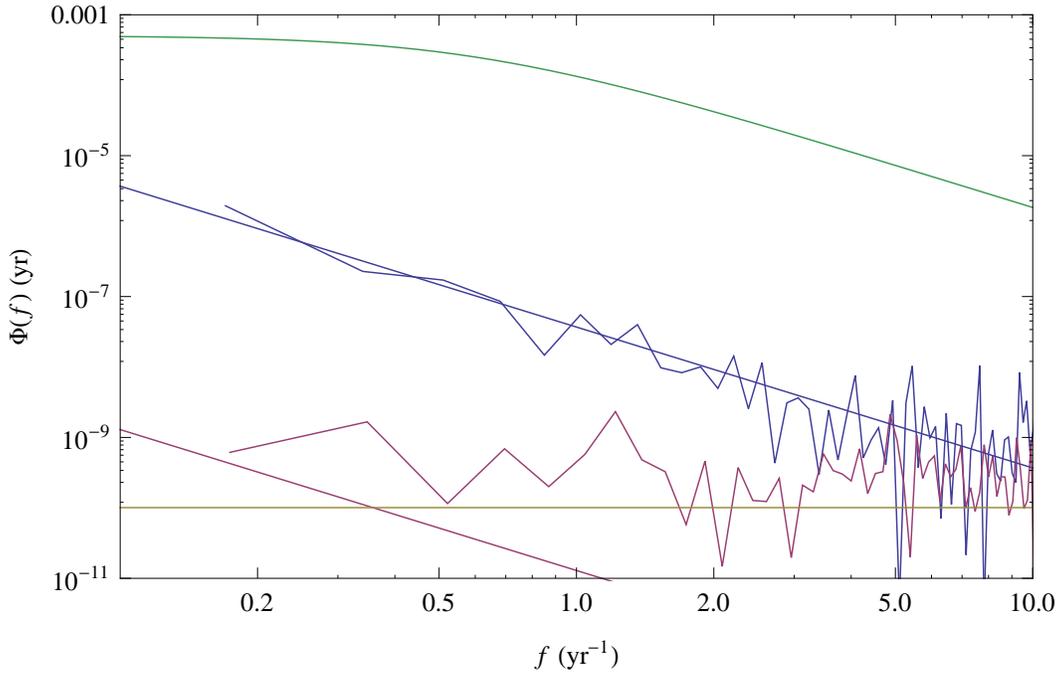} 
%
\end{center}
\caption{
Timing noise power spectral density $\Phi(f)$
(units: ${\rm yr}$)
as a function of Fourier frequency $f$
(units: ${\rm yr^{-1}}$)
for two representative millisecond pulsars,
PSR J1909$-$3744 (lower, purple curve)
and PSR J1939$+$2134 (upper, blue curve).
Overplotted are theoretical curves calculated from (\ref{eq:tmg12}) 
for $\lambda=3\times 10^{-2}$ and
$\eta(R^{-1})=(2\pi)^{-1/2}\varepsilon^{1/3} R^{-2/3} \gamma = 1\,{\rm yr^{-1}}$
(top, green curve),
$2\times 10^{-4}\,{\rm yr^{-1}}$,
(blue curve),
$7\times 10^{-8}\,{\rm yr^{-1}}$,
(purple curve),
$5\times 10^6\,{\rm yr^{-1}}$
(bottom, brown curve).
}
\label{fig:tmg1}
\end{figure}

Existing data already permit consequential tests of the theory.
As an example,
Figure \ref{fig:tmg1} displays the timing noise spectra of two
representative millisecond pulsars,
one quiet (PSR J1909$-$3744; lower, purple curve)
and one noisy (PSR J1939$+$2134; upper, blue curve).
The power spectral density $\Phi(f)$ (vertical axis) 
is plotted in units of yr;
we convert from the units of ${\rm yr}^3$ favored elsewhere
\citep{hob10}
by dividing by the spin period squared,
so that multiple objects can be compared meaningfully on the same plot.
The flat portions of the two zig-zag curves correspond to
white Gaussian noise arising from measurement errors and the 
ephemeris fitting process,
as well as possibly a component intrinsic to the pulsar.
The data are post-processed by jointly whitening the 
low-pass-filtered phase residuals and timing model by applying
a Cholesky transformation to the
covariance matrix to compensate for correlated noise
\citep{col11}.
The whitened correlations
arise chiefly from inadequate calibration of the raw observations
and imperfect correction for variations in the interstellar dispersion;
i.e.\ they are predominantly extrinsic
\citep{col11}.
The left-hand portion of the blue curve is genuine timing noise,
with a red spectrum below $f \lesssim 3\,{\rm yr^{-1}}$.
Overplotted are theoretical curves for $\lambda=3\times 10^{-2}$
and four values of
$\varepsilon^{1/3} R^{-2/3} \gamma$
specified in the caption.

A striking feature of Figure \ref{fig:tmg1} is that the predicted 
phase noise amplitude is high;
superfluid turbulence can perturb the rotational phase of the crust 
at an observable level.
The top (green) theoretical curve, 
plotted for $\eta(R^{-1})=1\,{\rm yr^{-1}}$,
sits well above the data.
Generally, at a particular observation frequency $f_0$,
the theoretical spectral power peaks for $\eta(R^{-1}) \approx 1.6 f_0$,
i.e.\ when the spectrum rolls over near $f_0$.
The theoretical peak amplitude,
$\Phi(f_0) \approx 2 \times 10^{-7} \lambda^{-2} f_0^{-1}$,
typically exceeds the observed spectral power by a wide margin
for traditional values of the crust's moment of inertia,
viz.\ $10^{-2} \lesssim \lambda \lesssim 0.5$
\citep{lyn00,lat07,van10,has12}.

In order to pull the theoretical curve below the observations,
the decorrelation frequency $\eta(R^{-1})$ must fall well below
or well above the observation band.
For slow decorrelation,
i.e.\ $\eta(R^{-1}) \ll f_0$, we obtain a red spectrum of the form
$\Phi(f) \propto f^{-2}$
within the observation band.
This scenario corresponds to the two middle, diagonal curves,
whose parameters are chosen to match the red noise signal measured
in PSR J1939$+$2134 
[blue curve; $\eta(R^{-1})=2\times 10^{-4}\,{\rm yr^{-1}}$]
and to lie underneath the white noise background measured
in PSR J1909$-$3744 to give an upper bound
[purple curve; $\eta(R^{-1})\leq 7\times 10^{-8}\,{\rm yr^{-1}}$].
The agreement with PSR J1939$+$2134 is excellent given the simplicity
of the model, and the inferred limits on 
$\varepsilon^{1/3} R^{-2/3} \gamma$ 
are reasonable for both objects,
as discussed below.
For fast decorrelation, i.e. $\eta(R^{-1}) \gg f_0$, 
we obtain a white spectrum 
$\Phi(f) \approx {\rm constant}$
within the observation band.
This scenario cannot explain the red noise in PSR J1939$+$2134,
which would then arise from a different physical process,
but it still constrains the turbulence parameters usefully:
the bottom (brown, horizontal) theoretical curve yields an approximate
lower bound
$\eta(R^{-1}) \geq 5 \times 10^6\,{\rm yr^{-1}}$
for both objects.
It is straightforward to compute the above bounds as functions of
the moment-of-inertia ratio $\lambda$.

What do constraints like those above imply for $\varepsilon$?
At a microscopic level, the power per unit enthalpy injected into the turbulence
by the relevant vortex or hydrodynamic instability
equals the growth rate of the instability multiplied by the square of 
its saturation velocity
\citep{per06a,and07}.
Numerical simulations of candidate instabilities
have not yet been conducted in the nonlinear regime under neutron star conditions,
so we relate $\varepsilon$ to observable quantities
by appealing to overall energy balance:
the power dissipated in the turbulence,
$\frac{4}{3} \pi R^3 \rho \varepsilon$,
equals the work done per unit time by the spin-down torque against pinning,
$(1-\lambda) I_0 \dot{\Omega} \Delta\Omega$,
leading to
\begin{equation}
 \varepsilon
 = 
 \frac{2}{5} (1-\lambda) R^2 \dot{\Omega} \Delta\Omega~,
 \label{eq:tmg15b}
\end{equation}
where 
$\dot{\Omega}$ is the spin-down rate,
and $\Delta\Omega$ is the steady-state angular velocity lag between the
condensate and charged electron-proton fluid maintained by pinning.
\footnote{
\citet{lan59} proposed $\varepsilon \propto (\Delta\Omega)^3$
for Navier-Stokes turbulence driven by a constant shear $\Delta\Omega$,
which produces a more energetic flow (and hence stronger timing noise) than
$\varepsilon \propto \dot{\Omega} \Delta\Omega$ for typical pulsar parameters.
Here we stick with the latter alternative to be conservative,
noting only that there is legitimate debate around what form of $\varepsilon$
suits the boundary conditions best;
see also \citet{mel10}.
}
Equation (\ref{eq:tmg15b}) shows that,
as the crust undergoes fluctuations in
$\dot{\Omega}$ and $\Delta\Omega$,
$\varepsilon$ fluctuates by a fractional amount of order
$\langle \delta\Omega_z^2 \rangle^{1/2} / \Omega
 = (I_{\rm c}\Omega)^{-1} G_{zz}(0)^{1/2}
 = 0.011 \lambda^{-1} (1-\lambda)^{1/3}
 (\dot{\Omega} \Delta\Omega/\Omega^3 )^{1/3} \gamma
 \lesssim 10^{-7} \lambda^{-1} \gamma \ll 1$
in a typical star,
justifying our neglect of the back reaction of the crust motion
on the velocity autocorrelation function of the superfluid turbulence
[equation (\ref{eq:tmg4})].

We can use (\ref{eq:tmg15b}) to convert an observational upper/lower bound 
on the decorrelation time-scale,
$\eta(R^{-1})_{\rm obs}$,
into an upper/lower bound on $\gamma$ via
\begin{equation}
 \gamma
 =
 2.3\times 10^{-3}
 (1-\lambda)^{-1/3}
 (
  {\dot{\Omega}} / {10^{-13}\,{\rm s^{-2}}}
 )^{-1/3}
 (
  {\Delta\Omega} / {1\,{\rm s^{-1}}}
 )^{-1/3}
 [
  {\eta(R^{-1})_{\rm obs}} / {1\,{\rm yr^{-1}}}
 ]~.
 \label{eq:tmg16}
\end{equation}
When applied to Figure \ref{fig:tmg1},
equation (\ref{eq:tmg16}) leads to one of two interesting conclusions.
If the decorrelation is slow, so that $\Phi(f)$ falls as $f^{-2}$
in the observation band,
we infer from the figure that $\gamma$ is small, with
$\gamma = 3\times 10^{-7} (\Delta\Omega/1\,{\rm s^{-1}})^{-1/3}$
for PSR J1939$+$2134 (blue curve) and
$\gamma \leq 3\times 10^{-10} (\Delta\Omega/1\,{\rm s^{-1}})^{-1/3}$
for PSR J1909$-$3744 (purple curve).
These values are consistent with other, independent evidence of 
strong but imperfect pinning.
On the other hand, if the decorrelation is fast,
so that $\Phi(f)$ is flat across the band,
we infer that $\gamma$ approaches unity,
consistent with no superfluidity.
In fact, the brown curve in Figure \ref{fig:tmg1} implies
$\gamma \gtrsim 8\times 10^{3} (\Delta\Omega/1\,{\rm s^{-1}})^{-1/3}$,
contradicting the physical requirement $\gamma\leq 1$.
Hence, for PSR J1939$+$2134 and PSR J1909$-$3744 specifically, the data and theory
together rule out fast decorrelation; even for $\gamma=1$,
the predicted $\Phi(f)$ exceeds the observations.
However, fast decorrelation remains a valid scenario in objects 
that spin down faster and/or are noisier than the pair in Figure \ref{fig:tmg1}.

The above analysis can be extended fruitfully to other objects. 
High-quality power spectral density curves
like those displayed in Figure \ref{fig:tmg1} are challenging to generate.
We plan to undertake a systematic analysis of more objects in the near future,
as more data flow out of
pulsar timing array projects searching for gravitational waves 
\citep{hob10,yar11,man12}
and timing noise experiments targeting young pulsars
\citep{zha12}.
\footnote{
R. Shannon, private communication.
}

\subsection{Amplitude versus spin-down rate
 \label{sec:tmg4b}}
The theory also predicts how timing noise varies in strength
across the pulsar population. 
One practical, unbiased measure is the power spectral density $\Phi(f_0)$ 
at some reference frequency, $f_0$, which is adequately sampled
in every pulsar under investigation
and avoids the Earth's orbital frequency and its harmonics.
Assuming fast decorrelation (red spectrum), 
equation (\ref{eq:tmg15}) implies
\begin{equation}
 \Phi(f_0)
 \propto
 f_0^{-2} \lambda^{-2}
 \dot{\Omega}^{1/3} (\Delta\Omega )^{1/3} \gamma~.
\label{eq:tmg17}
\end{equation}
The lag $\Delta\Omega$ is regulated by Magnus and pinning forces;
specifically it is 
set by the spacing of pinning sites, the pinning potential,
and the superfluid coherence length,
and therefore depends on thermodynamic variables like density and temperature
but not on the rotational state.
Reasonable estimates are 
$\Delta\Omega \sim 0.1 (B/10^{12}\,{\rm G}) \, {\rm rad\,s^{-1}}$
in the core, where $B$ is the magnetic field strength
\citep{lin12a},
and 
$\Delta\Omega \sim 0.4 \, {\rm rad\,s^{-1}}$
\citep{lin12b}
to
$\Delta\Omega \sim 0.1 \, {\rm rad\,s^{-1}}$
\citep{gri12}
in the inner crust.
Likewise, $\lambda$ and $\gamma$ are set by nuclear physics and
do not trend systematically with $\Omega$ and $\dot{\Omega}$.
They span a wider range than $\Delta\Omega$,
e.g. $10^{-2} \lesssim \lambda \lesssim 0.5$
\citep{lat07,van10}.
Smaller values in this range correspond to the solid crust plus
entrained neutrons, 
while larger values correspond to tight magnetic coupling between
the crust and core.
Equation (\ref{eq:tmg17}) therefore implies
\begin{equation}
 \Phi(f_0) \propto \dot{\Omega}^{1/3}~.
\label{eq:tmg18}
\end{equation}
Some scatter is expected around (\ref{eq:tmg18}),
because $\gamma$, $\lambda$, and $\Delta\Omega$
vary from one object to another;
on the other hand,
$\dot{\Omega}$ spans six decades across the isolated pulsar population,
so a trend should be discernible.
Measuring $\Phi(f)$ dependably is a demanding task, but progress 
has been made recently in the course of perfecting experiments
to detect gravitational radiation with pulsar timing arrays.
\footnote{
G. Hobbs, private communication.
}

In the time domain, the theory predicts how far the phase wanders 
stochastically from the underlying, deterministic ephemeris
over the observation time $T_{\rm obs}$.
The wandering is quantified
(i) cumulatively,
in terms of the cubic Taylor series term
$(12\pi)^{-1} \ddot{\Omega} T_{\rm obs}^3$
left over after subtracting $\Omega$ and $\dot{\Omega}$ from $\phi(t)$,
or (ii) progressively,
in terms of the root-mean-square phase residuals
$\sigma=\langle \delta\phi(t)^2 \rangle^{1/2}$
for $0\leq t \leq T_{\rm obs}$.
Time-domain tests are more ambiguous than a straight measurement of
the power spectral density, because there are many competing ways to
subtract polynomial and/or harmonic terms from the time series,
each introducing a degree of whitening that is difficult to quantify.
Still, despite the risk of ambiguity,
time-domain tests have certain advantages:
they are quick, they can be attempted on many objects with existing data,
and they are independent of frequency-domain tests, 
in the sense that they address the
nonstationary component of the timing noise, which is explicitly subtracted
to get $\Phi(f)$ in \S\ref{sec:tmg3}.

One popular measure of cumulative phase wandering is the dimensionless
stability statistic
$\sigma_z=(720)^{-1/2} \Omega^{-1} 
 \langle \ddot{\Omega}^2 \rangle^{1/2}
 T_{\rm obs}^2$
\citep{mat97}.
$\sigma_z$ is a generalized form of the Allan variance
(used to characterize the stability of man-made clocks)
computed from third-order differences in residuals
(to exclude fixed frequency drifts)
by fitting a cubic polynomial rather than explicit differencing
(because pulsars are monitored in irregular blocks of time separated by gaps)
\citep{mat97,hob10,sha10}.
Polynomial coefficients are not predicted directly
by the Kolmogorov theory in \S\ref{sec:tmg3}, 
but the root-mean-square amplitude of the cubic term can be estimated
statistically from
$\langle \delta\ddot{\Omega}(t)^2 \rangle^{1/2}
 =
 I_{\rm c}^{-1} \langle d^4G_{zz}/d\tau^4 \rangle_{\tau=0}^{1/2}$
\citep{kos02,mel10}, 
leading to the prediction
\begin{eqnarray}
 \sigma_z
 & = &
 1.97 \times 10^{-11}
 (\lambda / 0.03)^{-1} (1-\lambda)
 (\Delta\Omega / 1\,{\rm s^{-1}})
 (\gamma / 10^{-4})^3
 \nonumber
 \\
 & &
 \times
 ( \Omega / 2\pi \, {\rm s^{-1}} )^{-1}
 (\dot{\Omega}/10^{-13}\,{\rm s^{-2}})
 (T_{\rm obs}/ 10\,{\rm yr})^2~.
\label{eq:tmg19}
\end{eqnarray}

\begin{figure}
\begin{center}
\includegraphics[width=14cm,angle=0]{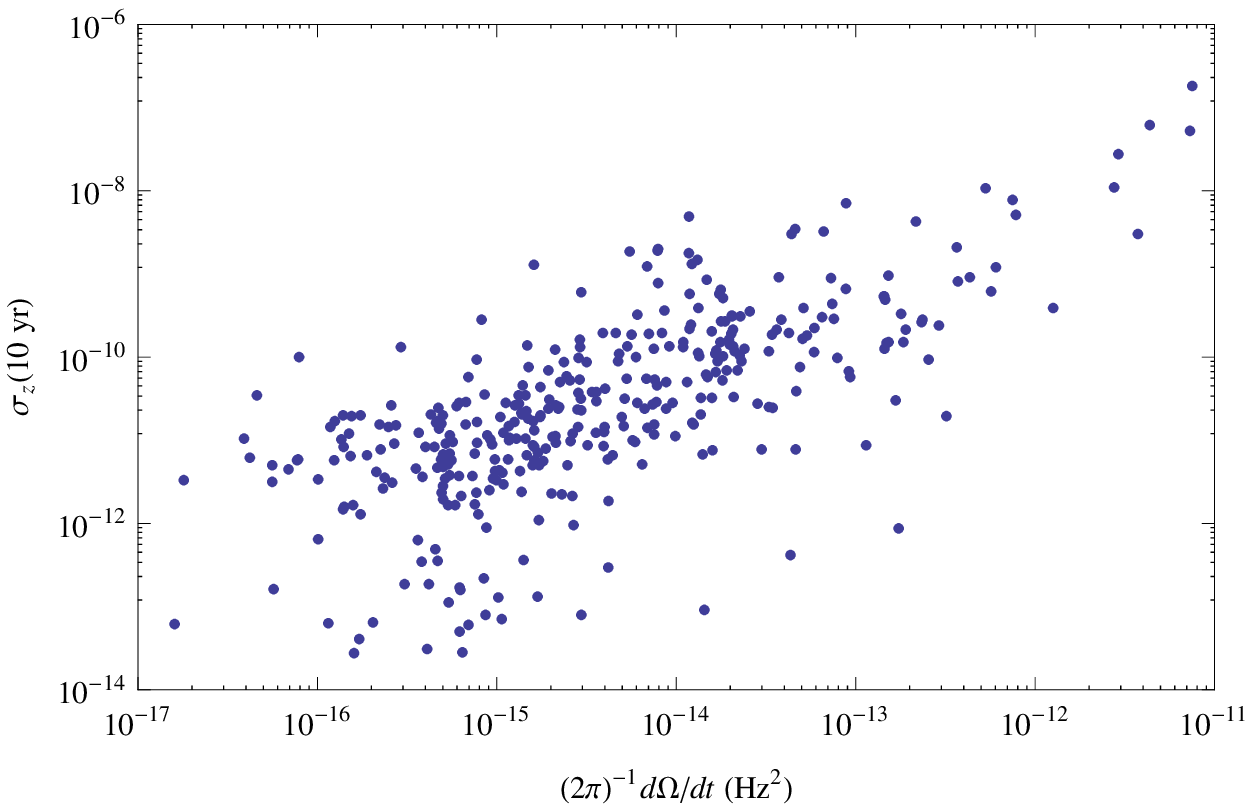} 
\includegraphics[width=14cm,angle=0]{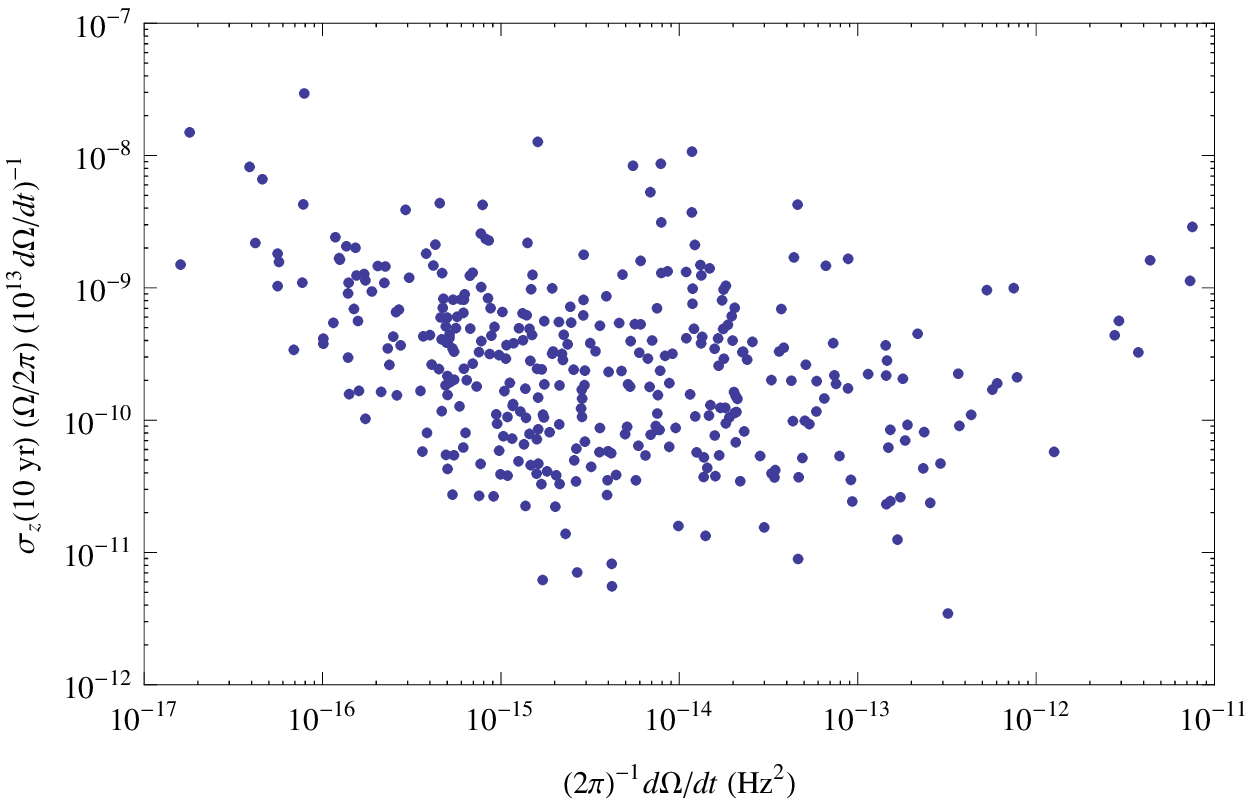} 
%
\end{center}
\caption{
Generalized Allan variance $\sigma_z(10\,{\rm yr})$
(dimensionless)
as a function of spin-down rate $\dot{\Omega}/(2\pi)$
(in units of ${\rm Hz^2})$
for the pulsar sample in \citet{hob10}.
({\em Top panel.}) Raw data.
({\em Bottom panel.}) Raw data multiplied by the factor
$(\Omega/2\pi\,{\rm s^{-1}}) (\dot{\Omega} / 10^{-13}\,{\rm s^{-2}})^{-1}$.
The theory predicts that the bottom panel should exhibit no trend.
}
\label{fig:tmg3}
\end{figure}

Figure \ref{fig:tmg3} displays 
$\sigma_z(10\,{\rm yr}) = \sigma_z(T_{\rm obs}={\rm 10\,yr})$ 
as a function of
$\dot{\Omega}$ for the pulsar sample analyzed by \citet{hob10}.
In the top panel, we plot the raw data.
In the bottom panel,
we plot the normalized quantity
$\sigma_z(10\,{\rm yr}) (\Omega/2\pi \,{\rm s^{-1}})
 (\dot{\Omega} / 10^{-13}\,{\rm s^{-2}})^{-1}$.
According to (\ref{eq:tmg19}),
the normalized $\sigma_z$ should be independent of 
$\Omega$, $\dot{\Omega}$, and $T_{\rm obs}$;
its scatter should reflect the scatter in the nuclear-related quantities 
$\lambda$, $\Delta\Omega$, and $\gamma$ across the pulsar population.
Do the data support this?
On balance, yes.
The raw $\sigma_z$ values span more than seven decades and display
a clear trend with $\dot{\Omega}$
(Pearson correlation coefficient 0.83).
\footnote{
Errors in $\dot{\Omega}$ leak into $\ddot{\Omega}$ to leading order,
when $\dot{\Omega}$ is subtracted from the ephemeris $\phi(t)$,
and hence may explain part of the trend in $\sigma_z$
versus $\dot{\Omega}$.
}
By contrast, the normalized data span four decades
and do not display a statistically significant trend with $\dot{\Omega}$
(Pearson correlation coefficient $-0.0052$),
as predicted by the theory.
This is encouraging, given how little is known about the precise
form of the temporal correlations in Kolmogorov turbulence even in
terrestrial experiments, let alone a neutron star superfluid.
The spread in the normalized data 
[11 out of 366 points in the bottom panel of Figure \ref{fig:tmg3}
lie between
$10^{-11}$ and $10^{-8}$ on the vertical axis]
is consistent with $\gamma$ varying moderately by a factor of $\sim 10$
across the population.
The rough proportionality between $\sigma_z$ and $\dot{\Omega}$ is 
also in accord with many previous studies
\citep{cor80,arz94,mat97,hob10,sha10};
for example, \citet{hob10} found 
$\sigma_z \propto \Omega^{-0.40} \dot{\Omega}^{0.75}$
with Pearson correlation coefficient $0.77$.
Similar conclusions hold for the stability parameter $\Delta_8$
\citep{arz94},
which expresses $\langle \ddot{\Omega}^2 \rangle^{1/2}$
for $T_{\rm obs} = 10^8\,{\rm s}$
in terms of the logarithm of a dimensional quantity
(essentially $\sigma_z T_{\rm obs}$).

In gravitational wave detection experiments with pulsar timing arrays,
it is customary to whiten the phase residuals by subtracting harmonically
related sinusoids until the red spectral component is nullified
\citep{hob10,yar11,man12}.
The whitened residuals contain uncorrelated instrumental noise
and intrinsic, flat-spectrum rotational irregularities in unknown proportions,
summed in quadrature.
Superfluid turbulence predicts the existence of a flat-spectrum
timing noise component of exactly this sort, namely
the low-frequency end of $\Phi(f)$ at $f\ll \eta(R^{-1})$,
which observations cannot resolve spectrally at present
for $\eta(R^{-1}) \lesssim 0.03\,{\rm yr^{-1}}$,
but which dominates the spectral power and cumulative
root-mean-square phase residual
$\sigma=\langle \delta\phi(T_{\rm obs})^2 \rangle^{1/2}$.
Upon setting $t=t'=T_{\rm obs}$ in (\ref{eq:tmg9c}),
we are left post-whitening with 
$\sigma\propto \lambda^{-1} [\eta(R^{-1})]^{1/2} T_{\rm obs}^{1/2}$
or, numerically,
\begin{eqnarray}
 \sigma
 & = &
 5.59 \times 10^{-2}
 (\lambda / 0.03)^{-1} (1-\lambda)^{1/6}
 (\Delta\Omega / 1\,{\rm s^{-1}})^{1/6}
 (\gamma / 10^{-4})^{1/2}
 \nonumber
 \\
 & &
 \times
 (\dot{\Omega}/10^{-13}\,{\rm s^{-2}})^{1/6}
 (T_{\rm obs}/ 10\,{\rm yr})^{1/2}~.
\label{eq:tmg20}
\end{eqnarray}
This is a striking result:
the intrinsic component of the whitened residuals
depends weakly on $\dot{\Omega}$, 
leaving the factor $\lambda^{-1} (\Delta\Omega)^{1/6} \gamma^{1/2}$ 
as the main source of variation across the pulsar population.

\begin{figure}
\begin{center}
\includegraphics[width=14cm,angle=0]{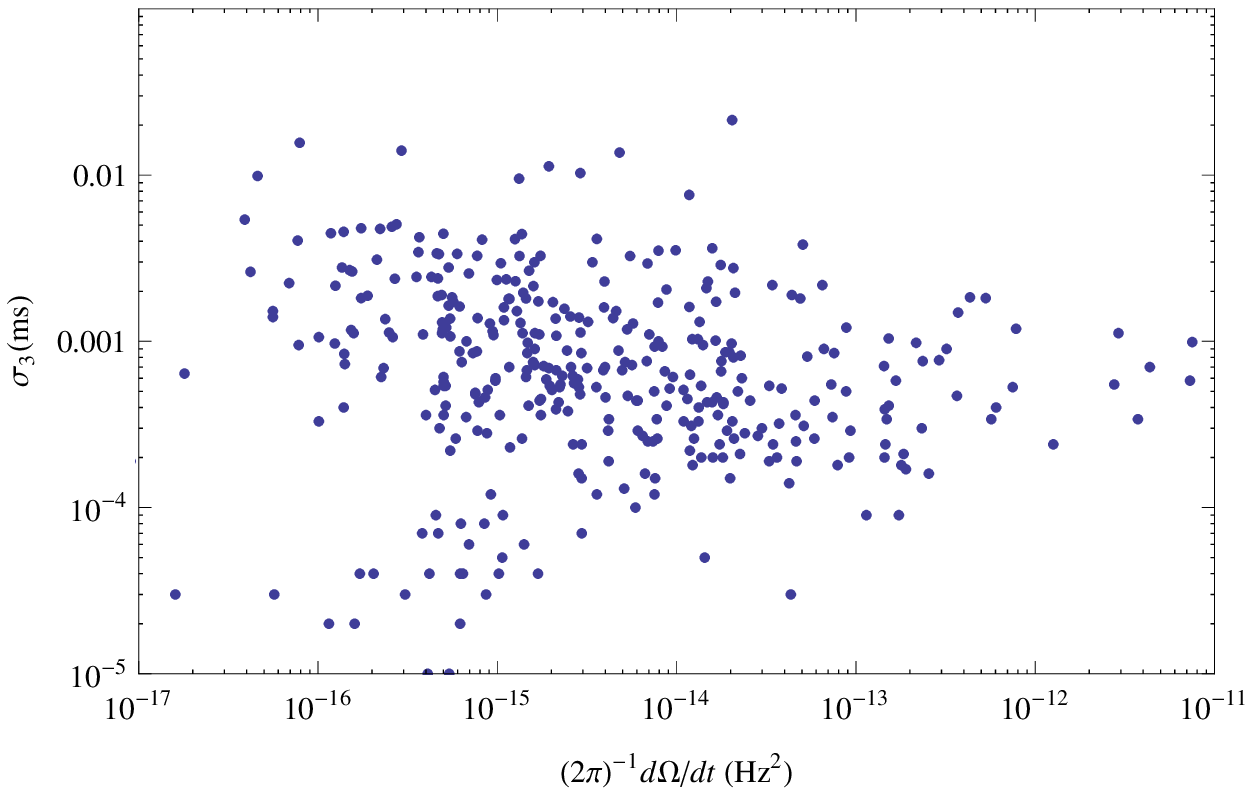} 
\includegraphics[width=14cm,angle=0]{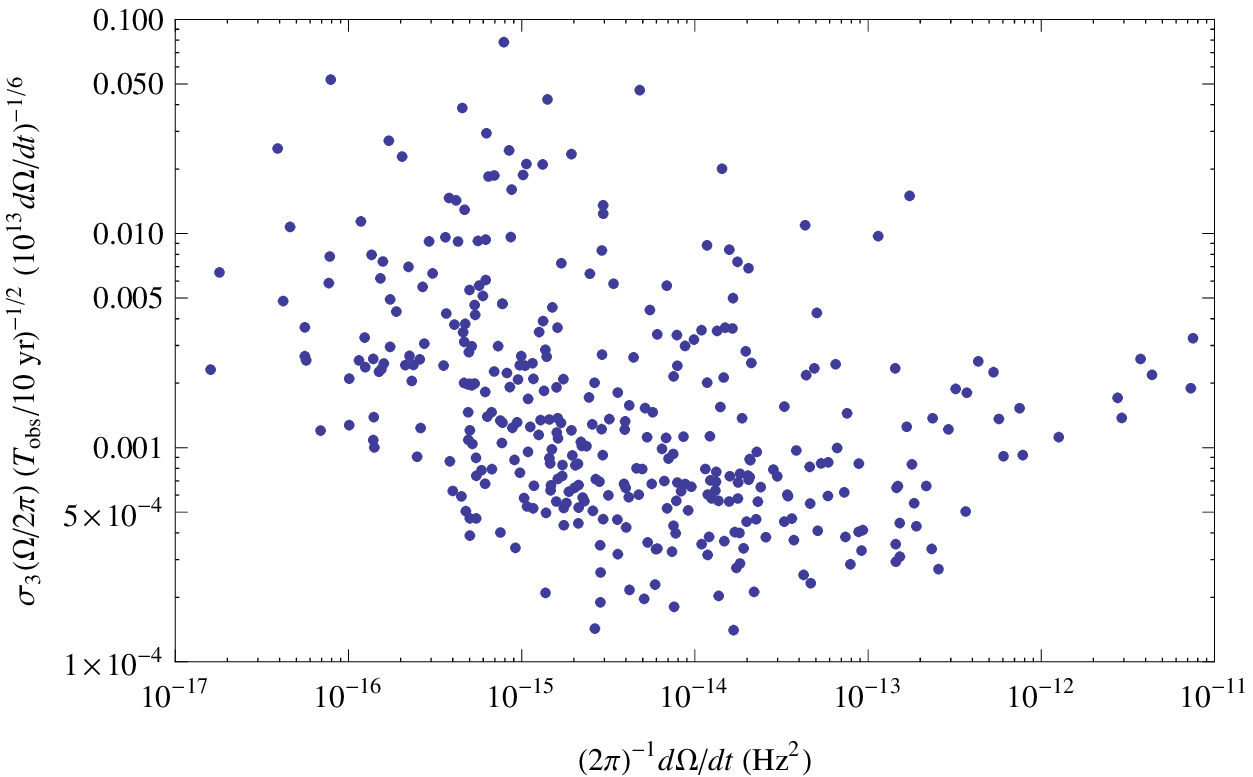} 
%
\end{center}
\caption{
Whitened root-mean-square phase residuals $\sigma_3$ after subtracting a
two-term $(\Omega,\dot{\Omega})$ ephemeris
as a function of spin-down rate 
$\dot{\Omega}/(2\pi)$ (in units of ${\rm Hz}^2$)
for the pulsar sample in \citet{hob10}.
({\em Top panel.})
Raw data in units of ms.
({\em Bottom panel.})
Raw data multiplied by the factor
$\Omega/(2\pi \, {\rm s^{-1}})
 (\dot{\Omega}/10^{-13}\,{\rm s^{-2}})^{-1/6}
 (T_{\rm obs}/ 10\,{\rm yr})^{-1/2}$.
The theory predicts that the bottom panel should exhibit no trend.
}
\label{fig:tmg4}
\end{figure}

Figure \ref{fig:tmg4} displays the whitened root-mean-square residuals
as a function of $\dot{\Omega}$ for the pulsar sample analyzed by \citet{hob10}.
In the top panel, we plot the raw data in units of ms,
i.e.\ the quantity labeled $\sigma_3$ in \citet{hob10}.
In the bottom panel,
we plot the normalized residuals
$\sigma_3 (\Omega/2\pi\,{\rm s^{-1}})
 (\dot{\Omega}/10^{-13}\,{\rm s^{-2}})^{-1/6}
 (T_{\rm obs}/ 10\,{\rm yr})^{-1/2}$,
converting $\sigma_3$ into a dimensionless quantity,
which can be compared directly with $\sigma$ in (\ref{eq:tmg20}).
It is difficult to disentangle the instrumental and intrinsic
components of $\sigma$ without further investigation
(e.g.\ altering the instrumental configuration).
Neither the raw nor the normalized data exhibit a trend with $\dot{\Omega}$
in Figure \ref{fig:tmg4}
(Pearson correlation coefficients $-0.047$ and $-0.033$ respectively),
in keeping with the prediction of (\ref{eq:tmg20}) but also with what
one expects if the noise is instrumental.
The normalized data span three decades,
making it unlikely that intrinsic noise dominates instrumental noise
in every object in Figure \ref{fig:tmg4};
otherwise, equation (\ref{eq:tmg20}) would imply
that $\lambda^{-1} (\Delta\Omega)^{1/6} \gamma^{1/2}$ spans three decades too,
which is conceivable but unlikely in the light of
independent empirical studies of glitch recovery time-scales
\citep{van10,esp11,yu12},
the time-averaged spin-up rate due to glitches
\citep{lyn00,esp11},
and nuclear physics calculations
\citep{lat07}.
\footnote{
The range covered by the normalized $\sigma_3$ does not change significantly,
when we exclude 
the 25 millisecond pulsars with periods shorter than $10\,{\rm ms}$,
whose residuals are systematically lower 
($\sigma_3 \leq 0.09\,{\rm ms}$),
and the pulsars which are known to glitch
\citep{esp11},
whose quasi-exponential recoveries may pollute $\sigma_3$.
}
On the other hand,
there is not enough evidence to support the opposite conclusion,
namely that instrumental noise dominates intrinsic noise in every object.

We check (\ref{eq:tmg20}) for consistency
by inferring limits on $\lambda$ from the data then asking
whether they are sensible on theoretical grounds.
At one somewhat unlikely extreme,
if every observed object is dominated by intrinsic noise,
the data imply $\sigma \leq 8\times 10^{-2}$
(upper envelope of the points in the bottom panel of Figure \ref{fig:tmg4})
and hence 
$(\lambda/0.03)^{-1} (\Delta\Omega/1\,{\rm s^{-1}})^{1/6} (\gamma/10^{-4})^{1/2}
 \leq 1.4$ from (\ref{eq:tmg20}).
This lower bound agrees well with independent empirical and theoretical studies 
\citep{lyn00,lat07,van10,esp11}
and is already astrophysically interesting.
Reducing the instrumental component of $\sigma_3$
will tighten the bound.
At the other extreme,
if all the observed noise is instrumental,
the data imply
$\sigma \leq 1\times 10^{-4}$
and hence
$(\lambda/0.03)^{-1} (\Delta\Omega/1\,{\rm s^{-1}})^{1/6} (\gamma/10^{-4})^{1/2}
 \leq 2\times 10^{-3}$,
requiring
$\gamma \lesssim 10^{-10}$.
It is intriguing to speculate whether future observations will reduce the
instrumental component of $\sigma_3$, 
or whether we are starting to see an intrinsic white noise floor.
Reducing the instrumental noise ultimately creates an opportunity to falsify
the turbulence model, at least in its present idealized form,
if it proves possible to push the measured $\sigma$ and hence the
inferred $\gamma$ well below a physically reasonable value,
after allowing for the ambiguities inherent in the whitening process.
Further theoretical work is required to determine from first principles
what the lower limit on $\gamma$ should be.

\section{Conclusion
 \label{sec:tmg5}}
In this paper, we calculate analytically the statistics of the
rotational phase fluctuations produced by superfluid turbulence
in a neutron star in terms of two fundamental parameters:
the noncondensate fraction of the moment of inertia, 
$\lambda=I_{\rm c}/I_0$,
and the decorrelation time-scale, $\eta(R^{-1})^{-1}$,
which depends on the steady-state angular velocity shear and the 
dynamical response time of the superfluid.
The calculation is idealized, in the sense that the turbulence
is assumed to obey the isotropic Kolmogorov law, without allowing for
the undoubtedly important but poorly understood effects of buoyant
stratification, fast rotation, two-component superfluidity,
and turbulent hydromagnetic stresses
\citep{mel10,sal10,mel12}.
Simple formulas are given for the autocorrelation function of the
phase residuals in
the time [equation (\ref{eq:tmg9c})] and frequency
[equation (\ref{eq:tmg12})] domains.
It is shown that the spectrum is red, 
consistent with radio pulsar timing data.
Simple recipes are also presented for how to extract
$\lambda$ and $\eta(R^{-1})$ from the half-power point of the spectrum
[equations (\ref{eq:tmg13}) and (\ref{eq:tmg14})]
or place a limit on their product from the $f^{-2}$ tail
[equation (\ref{eq:tmg15})].
Steeper tails can be accommodated within the theory by modifying
slightly the exponential temporal decorrelation function
in (\ref{eq:tmg4}),
a generalization that will be considered in future work.

The theory is applied to data from a representative group 
of ordinary and millisecond pulsars to illustrate in a preliminary
fashion how the theory can be tested; 
a full comparison will be undertaken in a future paper.
For the objects studied, the decorrelation frequency is bounded by
$\eta(R^{-1}) \lesssim 10^{-3} \, {\rm yr^{-1}}$ or
$\eta(R^{-1}) \gtrsim 10^{6} \, {\rm yr^{-1}}$,
and the pinning response parameter satisfies
$\gamma\lesssim 10^{-4}$ 
or (for a limited sub-class of objects) 
$\gamma\sim 1$, 
consistent with other work
\citep{lin12b,lin12a}.
Superfluidity enters the theory purely through $\gamma$;
one has $\gamma=1$ for Navier-Stokes turbulence
and $\gamma\ll 1$ for pinned superfluid turbulence.
Hence the preference for $\gamma\ll 1$ implied by the data 
amounts to indirect yet independent evidence 
for superfluidity in neutron stars and warrants further study.
Good agreement is obtained with popular measures of the
root-mean-square phase residuals like the Allan variance $\sigma_z$,
both with respect to the overall normalization and the spin-down trend.
We show that whitened phase residuals can be used to place
astrophysically interesting bounds on $\lambda^{-1} \gamma^{1/2}$.
The results may find practical application to experiments currently
under way to detect gravitational radiation with pulsar timing arrays
\citep{haa11,yar11,man12},
chiefly by clarifyng the relative strength of the reducible 
and irreducible components in timing noise.

Additional observational tests are needed,
starting with extending the preliminary tests in this paper 
to more objects.
Direct measurements of $\Phi(f)$ are the cleanest signature
of the stationary component of the red noise
but they also require the greatest effort.
Root-mean-square residuals carry time-integrated information 
about the low-frequency, high-power component,
which cannot be resolved spectrally with existing, multi-decade data sets.
The challenge is to construct a stable root-mean-square statistic,
which does not depend on how the ephemeris is subtracted,
as many authors have noted previously 
\citep{cor80b,arz94,mat97,hob10,sha10,zha12}.
Equation (\ref{eq:tmg9c}), 
which gives the phase autocorrelation function,
lends insight into what additional tests
are likely to be profitable. 
One approach is to study 
the angular velocity residuals instead of the phase residuals, 
since the former, unlike the latter, constitute a stationary time series,
with
$\langle \delta\Omega_i(t) \delta\Omega_j(t') \rangle \propto G_{ij}(\tau)$
via (\ref{eq:tmg7}).
\citet{bay99} constructed spectra for angular velocity residuals
by removing quadratic and cubic trends simultaneously from
pulse-frequency and time-of-arrival data.
They found scalings of the form $f^{-q}$, 
with $0.4\lesssim q \lesssim 2.4$,
in the tail of the spectra of four pulsars with anomalous braking indices,
PSR B0823$+$26, PSR B1706$-$16, PSR B1749$-$28, and PSR B2021$+$51,
but with error bars on $q$ of between $\pm 0.5$ and $\pm 1.4$,
i.e.\ consistent with $G_{ij}(f) \propto f^{-2}$ but inconclusive.
\footnote{
Magnetar torque spectra may be analysed too,
e.g.\ Figures 10 and 11 in \citet{woo02}.
Magnetic stresses change the character of the turbulence,
e.g.\ its effective dimensionality, a topic for future work.
}
It is worth testing, perhaps via Monte-Carlo simulations,
whether the advantage of stationarity enjoyed by 
$\langle \delta\Omega_i(t) \delta\Omega_j(t') \rangle$
outweighs the disadvantage of differentiating numerically
the $\phi(t)$ time series generated by the timing software.
Finally, whatever the technique, the theory can be tested by observing
for longer and extending the spectrum to lower frequencies, where it is
predicted to rise to
$\Phi(0)\approx 1.2\times 10^{-2} 
 (\lambda/0.03)^{-2} (1-\lambda)^{-1/3}
 ( \dot{\Omega} / 10^{-13}\,{\rm s^{-2}} )^{-1/3}
 ( \Delta\Omega / 1\,{\rm s^{-1}} )^{-1/3}
 (\gamma/10^{-4} )^{-1}
 \, {\rm yr}$.

Time-integrated braking indices also contain information about
timing noise
\citep{joh99,hob10}.
They have anomalous absolute values as large as $\sim 10^4$,
which manifestly do not describe magnetic dipole braking.
Evidence exists that pulsars younger than $\sim 10^5\,{\rm yr}$
have predominantly positive braking indices dominated by
glitch recoveries,
whereas the braking indices of pulsars older than $\sim 10^5\,{\rm yr}$ 
are positive or negative with roughly equal likelihood and reflect some
non-glitch, non-magnetic process, possibly superfluid turbulence
\citep{ura06,hob10}.
More work is needed to determine how to extract from (\ref{eq:tmg9c})
a time-integrated braking index,
which is directly comparable to the available data.

The theory presented in this paper can be extended in several ways.
First, an improved description of superfluid turbulence
is required:
on a local level,
to calculate $\gamma$ from simulations that
account for the pinning microphysics,
and on a global level,
to account for stratification, hydromagnetic stresses,
and multiple superfluid components,
which influence the Kolmogorov physics as well as $\gamma$.
Work is under way along these directions,
but the problem is formidable even under terrestrial conditions and
is unlikely to be solved soon
\citep{sal10}.
Turbulence itself alters transport coefficients like the viscosity, 
both macroscopically through mixing length physics and
microscopically through scattering in a vortex tangle.
Second,
off-axis torque fluctuations cause the rotation axis to precess
(cf.\ Chandler wobble),
with the angular displacement set by the dissipation physics
(cf.\ \S\ref{sec:tmg3}).
The rigid crust and corotating charged fluid are asymmetric
in general under the action of elastic and hydromagnetic stresses 
\citep{mel00,mas11},
the pinned superfluid vorticity induces gyroscopic precession 
on the time-scale $2\pi\lambda^{-1} \Omega^{-1}$
\citep{sha77},
and secular and/or stochastic torques do not necessarily average
to zero over many precession cycles
\citep{mel00}.
It is interesting to speculate whether the absence or presence
of precession explains the different types of timing noise observed
in individual pulsars,
characterized as phase, frequency, and torque noise in the literature
\citep{cor80,cor85},
and whether there is any correlation with pulse profile/polarization variations
\citep{osl11}.
Third, it is worth asking whether the theory in this paper can help
relate the physics of timing noise and glitches,
in pulsars where both phenomena are present.
For example, if the observed post-glitch recovery reflects the 
dynamics of the core superfluid,
temporarily decoupled from the crust by strong flux tube pinning
\citep{lin13},
then $\eta(R^{-1})^{-1}$ can be identified approximately with the
recovery time-scale 
\citep{van10},
and the normalization of $\Phi(f)$ is proportional to the
recovery time-scale through (\ref{eq:tmg12}),
a testable prediction.

Finally, we emphasize that there is compelling evidence that
timing noise is dominated by magnetospheric state switching in certain pulsars
\citep{lyn10},
and that this phenomenon is not incorporated in the theory presented here.

\appendix

\section{Angular momentum autocorrelation function
 \label{sec:tmgappa}}
The angular momentum autocorrelation function can be written
in the form
\begin{eqnarray}
 \langle \delta J_i(t) \delta J_j(t')^\ast \rangle
 & = &
 \rho^2 \epsilon_{ilm} \epsilon_{jpq}
 \int_0^R dr\,r^3 \int_0^R dr'\, (r')^3
 \int \frac{dk \, k^2}{(2\pi)^3}
 \nonumber 
 \\
 & & \times
 V P(k) \exp[- \eta(k) | \tau | ]
 K_{lmpq}(k,r,r')~,
\label{eq:tmgappa1}
\end{eqnarray}
with
\begin{equation}
 K_{lmpq}(k,r,r')
 =
 \int d^2{\bf \hat x} \int d^2{\bf \hat x}'
 \int d^2{\bf \hat k}
 \exp(i{\bf k}\cdot{\bf x}) \exp(-i{\bf k}'\cdot{\bf x}')
 {\hat x}_l {\hat x}_p'
 (\delta_{mq} - {\hat k}_m {\hat k}_q)~.
\label{eq:tmgappa2}
\end{equation}
One can easily prove the symmetry properties
$K_{lqpm}=K_{lmpq}$ and
$K_{pmlq}=K_{lmpq}^\ast$.
One also obtains
$K_{lmpq}=0$,
except when $l=p$ or $m=q$.
Furthermore the case $l=p=m=q$ does not survive the contraction
with $\epsilon_{ilm}\epsilon_{jpq}$, leaving
\begin{equation}
 K_{1212} = K_{1313} = K_{2121} = K_{2323} = K_{3131} = K_{3232}
 = 
 \frac{64\pi^3}{15} j_1(kr) j_1(kr')~,
\label{eq:tmgappa3}
\end{equation}
where $j_1$ denotes a spherical Bessel function of unit order
of the first kind.

The $k$ integral in (\ref{eq:tmgappa1}) runs from $k_{\rm s} R = 2\pi$ to
$k_{\rm d}R = R(8\varepsilon / 27\nu^3)^{1/4} \gg k_{\rm s}R$.
Hence over most of the range we have to a good approximation
\begin{eqnarray}
 \left[ \int_0^{kR} dx \, x^3 j_1(x) \right]^2
 & = &
 [ 3 k R \cos kR + (k^2 R^2 - 3) \sin kR ]^2
\label{eq:tmgappa4}
 \\
 & \approx &
 \frac{9}{2} k^2 R^2 + \frac{1}{2} (k^2 R^2 - 3)^2~,
\label{eq:tmgappa5}
\end{eqnarray}
plus fast oscillations proportional to $\cos(2kR)$ and $\sin(2kR)$,
which integrate almost to zero over $k$
\citep{mel10}.
Equation (\ref{eq:tmg7}) thence follows directly.

\acknowledgments
The authors thank V. Ravi, G. Hobbs, W. Coles, R. Shannon, and
R. Manchester for stimulating discussions
and for preparing the observational data plotted in the figures.
The power spectral density curves for PSR J1909$-$3744 and
PSR J1939$+$2134 in Figure \ref{fig:tmg1} are provided by the
Parkes Pulsar Timing Array Project and have been published
in Figure 11 of \citet{man12}.
This research was supported by the Australian Research Council and
the USA National Science Foundation.

\bibliographystyle{mn2e}
\bibliography{glitchstat}

\end{document}